\documentstyle[aps,preprint,epsf]{revtex}
\begin{document}
\title{Testing the symmetrization postulate of quantum mechanics\\
and the spin-statistics connection} 
\author{Guglielmo M. Tino}
\address{Dipartimento di Scienze, Universit\`a di Napoli "Federico II" 
and INFM,\\ 
Complesso Universitario di Monte S. Angelo, Via Cintia, I-80126 Napoli, Italy} 
\date{\today} 
\maketitle 
\medskip
\begin{abstract}
Recent experimental tests of the symmetrization 
postulate of quantum mechanics are discussed. 
It is shown that in a strict sense
these experiments cannot test the validity of the symmetrization postulate,
but in most cases do test the spin-statistics connection.
An experiment is proposed that would allow to search for possible
violations of the symmetrization postulate.
\end{abstract}
\pacs{05.30.Jp, 33.20.-t, 42.62.Fi} 
\narrowtext 

\section{Introduction}
Several experiments have been reported that assert to test the symmetrization
postulate of quantum mechanics and/or the spin-statistics connection.
The symmetrization postulate and the spin-statistics connection are indeed 
amongst the
fundamental tenets of quantum mechanics.  The symmetrization postulate 
establishes
that in a system containing identical particles the only possible states are
either all symmetrical or all antisymmetrical with respect to permutations of
the particles.  In the first case, the particles are called bosons and follow
Bose-Einstein statistics; in the second case they are called fermions and follow
Fermi-Dirac statistics.  
Experiments
indicate that particles with integer values of spin are bosons, while
particles with odd-half-integer spin are fermions.
The reason why only symmetric and
antisymmetric states seem to occurr in nature and the connection with the spin
of the particles has been a puzzle since the early days of quantum mechanics 
\cite{puzzle}.
The spin-statistics theorem, proved by W. Pauli \cite{Pauli40} 
from the basic principles of quantum field theory and special relativity
such as the
requirement of local commutativity of observables, states
that, given the
choice between Bose and Fermi statistics, integer-spin particles must obey
Bose statistics and odd-half-integer-spin particles must obey Fermi statistics.
Proofs of the spin-statistics theorem
are reviewed in \cite{Duck9798}. 

Quantum mechanics would nevertheless 
allow also symmetries different from those
imposed by the symmetrization postulate, 
and theories have been developed allowing for small
deviations from conventional statistics which might have been masked
in the experiments performed so far. It is worth noting that no theory
so far predicts the possibility of observing a violation in a particular 
system or in some specific condition. Consistent theories can
be formulated, however, 
which would lead to different symmetry properties. Experiments
are needed then to discriminate between these theories imposing constraints
which are the more stringent the higher is the experimental precision.

In this paper, after a brief account of the theoretical background,
a discussion of experimental tests 
is presented, 
trying to clarify in a consistent manner their meaning. It is shown 
that most of the recent experiments should be considered
as testing the spin-statistics connection
rather than the validity of the symmetrization postulate. 
An experiment to search for violations of the symmetrization
postulate for nuclei is proposed.

\section{Theoretical background}

It is not the purpose of this paper to discuss in detail
the theories which lead to symmetry properties different from the
ones which are peculiar to bosons and fermions. 
A general survey can be found in \cite{Greenberg94}.
It is of interest, however, to put in evidence the possibility of  
theories allowing for small deviations from the usual symmetry relations,
whose search is the subject of the experiments discussed in this paper.
Such deviations can be expressed as a different symmetry of the state
under particle exchange or, in Fock-space representation, as a deformation 
of the algebra of the creation and annihilation operators.

A statistics intermediate between  
Bose and Fermi cases was first proposed in
\cite{Gentile40} considering the possibility that at most $n$ identical 
particles could occupy the same quantum state. 
This idea led to a generalized field theory \cite{Green53},
called parastatistics, in which the field operators obey
trilinear commutation relations instead
of the usual bilinear relations. These theories predict, however, gross
violations of statistics which are immediately 
excluded by experimental evidence.

The possibility of a continuous interpolation between bosonic 
and fermionic behaviours is given by "quons" \cite{Greenberg91}. 
The commutation and anticommutation relations are replaced by 
generalized bilinear commutation relations depending on
a parameter $q$ ($q$-mutators):
\begin{equation}
a_k a^+ _l - q a^+ _l a_k = \delta _{kl}, ~~~~~~ -1 \leq q \leq 1
\end{equation}
with the vacuum condition $a_k|0\rangle = 0$. As $q$ varies between
-1 and 1, the symmetry changes continuously from the completely
antisymmetric case (fermions) to the completely symmetric case
(bosons). It has been shown that this interpolation 
preserves positivity of norms and the non-relativistic
form of locality \cite{Greenberg92}. Other aspects are still
doubtful such as the possibility of accounting for local observables in 
a relativistic theory or for the existence of antiparticles.

Statistics 
other than Fermi and Bose have also been investigated
for one- and two- dimensional systems \cite{Leinaas77} and in connection
with anyon high-temperature superconducting systems \cite{Wilczek90}. 

Two important points must be noted to avoid confusion in the interpretation
of experimental tests. 
The first is that 
in a system including two identical
particles, only two "entangled" states are possible, namely 
a symmetric and an antisymmetric state:
\begin{eqnarray} 
\Psi_{S}(1,2) = \frac{1}{\sqrt{2}} 
[ \psi_{a}(1) \psi_{b}(2) + \psi_{b}(1) \psi_{a}(2)] \\ \nonumber
\Psi_{A}(1,2) = \frac{1}{\sqrt{2}} 
[ \psi_{a}(1) \psi_{b}(2) - \psi_{b}(1) \psi_{a}(2)]
\end{eqnarray}
In this case, the symmetrization postulate does not play
any role, the question being only the connection between the spin 
of the particles and the symmetry of the two-particle state.  
This is not the case for systems including more than two
identical particles, where mixed-symmetry states are possible. 
The second important remark is that,
as was pointed out in \cite{Amado80}, in the framework 
of ordinary quantum mechanics transitions between states of different
permutation symmetry cannot take place. This "superselection rule" is
a rigorous selection rule holding also in the presence of perturbations
such as collisions or external fields. It is not possible then
to consider states given by a coherent superposition of states
of different permutation symmetry. The system must be described
as an incoherent mixture which is represented by a density matrix.
In the case of two particles, for example,
the two-particles density matrix 
for small violations of Bose statistics 
is:
\begin{equation} 
\rho_2\:=\:(1-\frac{1}{2}\beta^{2})\rho_s\:+\:\frac{1}{2}\beta^{2}\rho_a.
\end{equation} 
\noindent 
where $\rho_{s(a)}$ is the symmetric (antisymmetric) two-particle density 
matrix.
Neglecting these important points led to a wrong interpretation of
some of the experiments. 

Using the notation
adopted in the literature, in the following $\beta^{2}/2$ indicates
the "symmetry-violation" parameter. Its real meaning needs to be specified,
however, for the particular physical system 
and the theoretical model considered. In \cite{Greenberg91}, for example,
it was shown that in the frame of a $q$-mutator theory, the value of
$\beta^{2}/2$ can be related to the value of the $q$ parameter.

\section{Experiments}

In this section, experiments performed to search for violations of the 
symmetrization postulate and/or of the spin-statistics connection are 
discussed. The focus is on recent experiments
and, in particular, on experiments on integer-spin nuclei in molecules.
For a review of the work before 1989, the reader is referred to
\cite{Greenberg89}. As already mentioned, 
some of the initial experiments
suffered from a misunderstanding of the constraints imposed by the
"superselection rule". It is shown here that also the 
interpretation given for 
some recent experiments is not correct.
In fact, although the papers published on this subject
usually present their "null results" as a confirmation of the 
validity of the symmetrization postulate, 
most of them should be considered as tests of the spin-statistics
connection. The argument is simple: as discussed above, 
in a two-particle system, only symmetric
or antisymmetric states are possible. The symmetrization
postulate does not play any role in this case. Experiments 
involving only two identical particles should then
be considered as tests of the spin-statistics connection rather than
tests of the symmetrization postulate.
An experiment that would allow a genuine test of the 
symmetrization postulate is proposed in the following.

\subsection{Experiments on half-integer-spin particles.} 

A few experiments have been performed 
to test the validity of the Pauli principle for half-integer-spin particles.
In particular, a high precision test on electrons was performed in
\cite{Ramberg90} by running a current through a
copper bar and searching for the X-rays that would be emitted
if some of the electrons introduced in the sample 
were captured by a copper atom and cascaded
down to the 1S state, which is already filled with two electrons. 
No signal was found and this was interpreted as giving a limit
$\beta ^2/2 \leq 1.7\times 10^{-26}$ to the probability
that a new electron introduced into copper would form a mixed-symmetry 
state with respect to the other electrons already
present in the copper sample. The large number of electrons
in the system was important to reach such a high sensitivity but,
on the other hand, makes the interpretation of the result
more complicated. Conclusions may change depending on whether
we consider the symmetry of the system composed by the injected electron
plus the electrons already present in the copper bar, or we consider
a model in which the electron collides with a copper atom and is
captured.

A simpler two-electron system was investigated in \cite{Deilamian95}.
A spectroscopic test was performed
on helium atoms, searching for a transition 
involving the permutation symmetric
$1s2s~^{1}S_{0}$ state. An upper bound $\beta ^2/2 \leq 5 \times10^{- 6}$ 
was set to a violation of the Pauli principle. In spite of the lower
sensitivity, the interpretation of this result is simpler. 
Since only two identical particles are involved,
this must considered as a test of the spin-statistics connection. 
Doubts can be raised, however, about what would be the chemical stability
of "paronic" atoms in ordinary samples.
In \cite{Deilamian95}, this was taken into account by having the atoms ionize
and recombine in a discharge before entering the detection region.

\subsection{Experiments on integer-spin particles.}

No accurate test for integer-spin particles
had been reported until recently. 
This is due to the fact that while
there are several systems in which a violation of the Pauli exclusion 
principle would be
detected as a signal on a zero background, the effect of a small
violation for particles following Bose-Einstein statistics would
usually manifest itself as a small
change in the properties of a many-particle system. This obviously
limits the achievable accuracy.
In \cite{Greenberg89}, a
bound to a possible violation of the generalized Bose statistics
for pions was inferred considering 
the $K_{L} \rightarrow  \pi^{+}\pi^{-}$ decay, 
which is usually considered as due to CP violation. A limit
of $\beta ^2/2 \leq 10^{-6}$ was obtained.\\

{\it Photons.}
 
Several papers have been published recently reporting
or proposing experiments to set a limit 
to possible violations of Bose statistics for photons
\cite{Fivel91,Manko95,Ignatiev96,DeMille99,Gerry97,Greenberg99}.
The fundamental nature of the photon and its peculiar properties
makes it very interesting to investigate this particle
in this context. 
It is hard, however, to find an experiment that would give 
a direct evidence of a violation with a significant sensitivity.
This is one case, in fact, in which a small deviation from normal
statistics would usually produce only a small signal over a large background.
An attempt to set a limit to a possible violation
of Bose statistics was made in \cite{Fivel91}, 
based on light intensities attainable
in laser systems.
In \cite{Manko95}, a possible dependence of the 
frequency of light on its intensity was searched for. This effect is expected
if a $q$-nonlinearity is introduced in the description of the electromagnetic
field. Since nonlinearities in the commutation relations give rise
to mixed-symmetry states, this experiment could also be reinterpreted
as a search for a violation of the symmetrization postulate
for photons. The connection is not straightforward though and was not
pursued in the paper.

In \cite{Ignatiev96}, the experimental upper limit on the 
two-gamma decay of $Z$-boson, $Z \rightarrow \gamma\gamma$, was used
to establish an upper bound $\beta ^2/2 \leq 10^{-2}$
on a possible small violation
of the exchange-symmetry for a system of two photons. The same
idea, based on what is called Landau-Yang theorem, was exploited
in \cite{DeMille99} to improve the limit by searching for 
the forbidden
$J=0 \rightarrow J'=1$ transitions in atoms excited by two photons 
of the same energy. A limit of $\beta ^2/2 \leq 7 \times 10^{-8}$
was set on the probability that two photons are in an 
exchange-antisymmetric state.
A different approach was followed in \cite{Greenberg99}. A 
very tight bound to a violation of statistics
for photons was inferred
considering photons and electrons as coupled "quons" and relating 
the bound for photons to that obtained in \cite{Ramberg90} for electrons.
Although this argument is indirect and model-dependent, it is very
interesting and it could also be extended to other particles.\\

{\it Nuclei in molecules.}

In 1931, Ehrenfest and Oppenheimer showed that 
a composite system of fermions is a boson or a fermion
depending on whether it is made of an even or an odd number
of fermions \cite{Ehrenfest31}. Considering the total 
angular momentum resulting from the constituents angular momenta,
an extension of the spin-statistics theorem to composite systems is
obtained. For this argument to be valid, it is necessary
that the interaction between the composite particles 
is negligible compared to the internal excitation energy
so that the internal structure can be neglected and
the system can be considered as a single particle.
This has a striking demonstration in recent experiments
on Bose-Einstein condensation of atoms \cite{Tino99}: 
the phase transition is observed only for the isotopes
of the atoms for which the sum of the number of
protons, neutrons and electrons gives an even number, that
is when the number of neutrons in the nucleus is even.
The same argument applies to the symmetry of systems including
identical nuclei.
In \cite{Hilborn90,Tino94}, it was proposed that a high-sensitivity
investigation of the spectra of molecules containing identical spin-0 nuclei
would allow to search for violations of Bose statistics for the nuclei. The
basic idea of these experiments is analogous to the one underlying the
experiment on electrons in the helium atom \cite{Deilamian95}. This represents
indeed a rare case in which a violation of the spin-statistics connection for
integer-spin particles would be detected on a virtually zero background with a
sensitivity comparable to the one achieved in experiments on fermions.
Let us consider a molecule containing two identical spin-0 nuclei as,
for example, $^{16}$O$_{2}$.
According to the Born-Oppenheimer approximation and neglecting
the coupling of the nuclear spin with the rest of
the molecule (which is not important
for these experiments since the spin of the nuclei is
zero), the total wave function $\psi_{t}$ can be written in the form
\begin{equation}
\psi_{t} = \psi_{e}\psi_{v}\psi_{r}\psi_{n}
\end{equation}
where $\psi_{e}$, $\psi_{v}$ and $\psi_{r}$
are the electronic, vibrational and rotational functions, respectively,
and $\psi_{n}$ is the nuclear spin function.
For integer-spin nuclei, the total wavefunction
$\psi_{t}$
must be symmetric in the exchange of two nuclei. 
Since the nuclear spin is zero, $\psi_{n}$ is 
obviously symmetric. 
The vibrational wave function $\psi_{v}$ is also unaltered
in the exchange of the nuclei because it depends only on the magnitude of the
internuclear distance. Since the total wavefunction $\psi_{t}$
must be symmetric,
only the states corresponding to even (odd) rotational quantum numbers are
allowed if $\psi_{e}$ is symmetric (antisymmetric) \cite{Herzberg50}.

In a first series of experiments \cite{Tino96,deAngelis96,Hilborn96}, 
the spectrum of the $^{16}$O$_{2}$ molecule was investigated 
searching for transitions between states which are antisymmetric
under the exchange of the two nuclei. 
An upper bound $\beta^2/2\leq$ 5$\times10^{-7}$ was set in \cite{deAngelis96}, 
and a similar, slightly less accurate,
result was obtained in \cite{Hilborn96}. Experiments on $^{16}$O$_{2}$
were later reported in \cite{Naus97,Gianfrani99}. In \cite{Gianfrani99},
a limit $\beta^2/2\leq$ 5$\times10^{-8}$ was obtained. 

In \cite{Modugno98} a new accurate test on $^{16}$O nuclei was
performed by investigating the vibrational spectrum 
of the $^{12}$C$^{16}$O$_2$ molecule. The CO$_2$ molecule 
has the same symmetry properties of O$_2$ but, since it
is triatomic, it has strong active vibrational bands in the infrared which are
lacking in O$_2$ spectra.  In particular, the intensity of the
combination band 12$^0$1 - 00$^0$0 around 2~$\mu$m 
investigated in \cite{Modugno98} is more than two
orders of magnitude larger than the electronic transitions of oxygen previously
investigated, and therefore the absorbance of a given population is
correspondingly larger. This results in an increased detection sensitivity. 
Since the ground electronic wavefunction
of the CO$_2$ molecule is symmetric in the exchange of the two $^{16}$O nuclei,
the rotational wavefunction in
the ground vibrational state must be symmetric and only even values for the
rotational quantum number J are allowed.
The R-branch
investigated in \cite{Modugno98} should then 
be composed only of R(2J) transitions.
A bound of
$\beta^2/2\leq$(2.1$\pm$0.7)$\times$ 10$^{-9}$ to the relative 
population of the forbidden states was deduced in this work.  
This experiment gives at present the most stringent test 
of the  spin-statistics
connection for $^{16}$O nuclei.
Further improvements are possible:  the sensitivity
could be increased by reducing the noise to the quantum level, by increasing the
absorption pathlength, or by selecting stronger transitions.  The fundamental
vibrational band of CO$_2$ around 4.3~$\mu$m, which is at least a factor of
thousand stronger than the one at 2~$\mu$m, could be investigated. 
A dramatic increase in the detection sensitivity could be
obtained by a resonant-ionization-spectroscopy detection scheme.
A similar test can be performed on other
spin-0 nuclei; a straightforward extension is the investigation of 
the spectrum of $^{18}$O$_2$ \cite{Gagliardi97}.  

An interesting prospect is
the investigation of spectra of molecules containing more than two identical
nuclei. As shown above, most of the experiments performed so far
involve only two identical particles so that they provide 
a test of the spin-statistics connection. 
In order to search for possible violations of the symmetrization postulate
of quantum mechanics, it is now clear that systems including
more than two identical particles should be considered. 
In this case, indeed,
more complex symmetries are possible and the Young diagrams 
do not reduce to the trivial completely symmetric
and completely antisymmetric cases. 
On the other hand, systems composed of very large numbers of 
identical particles can complicate the interpretation of
experimental results.
Molecules offer the
possibility of investigating more and more complex structures and,
as for the experiments performed on O$_{2}$ and CO$_{2}$, 
high detection sensitivity can be achieved by laser spectroscopy 
methods. 
A good candidate for this experiment is OsO$_4$, a highly symmetric 
molecule with four identical spin-0 nuclei.
In spite of the higher complexity of the spectrum with respect to 
the simpler molecules investigated previously, high 
resolution and high sensitivity
spectroscopy schemes have been developed, especially in the region around 
10 $\mu m$ which is of metrological interest. The good knowledge of 
molecular parameters makes it simpler to find the position of the relevant 
transitions and to separate them from spurious signals. In particular,
transition frequencies can be singled out that would represent 
a signature of a violation of the symmetrization postulate. 
The expected detection sensitivity is expected to be comparable to that
achieved in previous tests on molecules.

\section{Conclusions}

Several experiments confirm the validity of the spin-statistics connection
for various types of particles to a high level of accuracy. 
This provides a proof of the general formalism of quantum mechanics
and can be extended also to different particles.
In a recent paper \cite{Greenberg99bis}, 
the results obtained in \cite{Modugno98} for nuclei
was used to set a bound on possible violations 
of the Pauli exclusion principle for nucleons and for quarks.
On the other hand, not many experiments have really tested 
the validity of the symmetrization postulate of quantum mechanics
with high precision. A possible experiment on nuclei in molecules
was proposed in this paper which should allow to test the symmetrization 
postulate with an accuracy comparable to that achieved in previous
tests of the spin-statistics connection.\\

The author acknowledges useful discussions with G. Modugno
and thanks R.C. Hilborn for providing preprints of his
recent papers.

\end{document}